\documentclass[12pt]{article}
\usepackage{amsthm,amsmath}
\providecommand{\mcal}{\ensuremath{\mathcal}}

\newcommand{\transker}[1][\opr{T}]{\ensuremath{\left.\left<q+\frac{v}{2}\right|#1\left|q-\frac{v}{2}\right>\right.}}
\newcommand{\kernel}[1]{\ensuremath{\left.\left<q\right|#1\left|q'\right>\right.}}
\newcommand{\kernell}[1]{\ensuremath{\left.\left<q'\right|#1\left|q\right>\right.}}
\newcommand{\tkernel}[1]{\ensuremath{\left.\left<t_a(q)\right|#1\left|t_a(q')\right>\right.}}

\newcommand{\dirac}[2]{\ensuremath{\left<#1\left|#2\right>\right.}}
\newcommand{\opr}[1]{\ensuremath{\hat{\mathbf{\mathsf{#1}}}}}

\newcommand{\abs}[1]{\ensuremath{\left|#1\right|}}

\newcommand{\sign}[1]{\ensuremath{\mbox{sgn}\left(#1\right)}}

\begin{document}
\title{\textbf{Quantum-Classical Correspondence of Dynamical Observables, Quantization and  the Time of Arrival Correspondence Problem}\thanks{Invited talk at the ICQO 2000, Raubichi, Belarus, May 28-31, 2000, \textsl{Opt. \& Specs} \textbf{91} (2001) 429.}}
\author{Eric A. Galapon \thanks{email: egalapon@nip.upd.edu.ph}\\Theoretical Physics Group, National Institute of Physics\\University of the Philippines, Diliman Quezon City, 1101 Philippines}
\maketitle
\begin{abstract}
We raise the problem of constructing quantum observables that have classical counterparts without quantization. Specifically we seek to define and motivate a solution to the quantum-classical correspondence problem independent from quantization and discuss the general insufficiency of prescriptive quantization, particularly the Weyl quantization. We demonstrate our points by constructing time of arrival operators without quantization and from these recover their classical counterparts.
\end{abstract}


\section{Introduction}\label{intro}

It is generally believed that classical mechanics is  the
contraction of quantum mechanics in some appropriate limit of
vanishing $\hbar$ \cite{zhang}.  Thus in principle every
classical observable---e.g. a real-valued function in the phase
space---is the contraction of some quantum observable. However,
quantum observables are generally constructed by the quantization
of classical observables, the mapping of the algebra of functions
in the phase space to the algebra of self-adjoint operators
acting in some Hilbert space \cite{quanti}. Obviously this
introduces circularity when one invokes the correspondence
principle. This is unsatisfactory if quantum mechanics were to be
internally coherent and autonomous from classical mechanics.

This glaring unsatisfactory state of quantum affair has  been
ignored because quantization has been successful in most of its
practical applications in the first place. One can, for example,
mention its convincing accuracy in predicting atomic spectra by
the quantization of the classical Hamiltonian of the classically
interacting charged particles. This success has encouraged the
development of numerous theories in  quantization, such as
geometric and deformation quantizations among others. It is
well-known though that there are obstructions to quantization
\cite{gotay}. For example, in flat Euclidean space, Groenwold and
van Hove \cite{grovan} have shown that there exists no
quantization of all classical observables consistent with
Schrodinger's quantization scheme. On the other hand,
prescriptive mapping of classical observables to quantum
observables, such as Weyl prescription, is known to be generally
inconsistent with Dirac's ``Poisson bracket-commutator
correspondence".

The existence of obstructions to quantization inevitably limits
the class of dynamical observables that can be consistently
quantized. This sobering handicap and the circularity of
quantization force upon us to re-examine the problem of
constructing quantum observables corresponding to the classical
ones. In this paper we raise the problem of constructing quantum
observables that have classical counterparts  without
quantization. Specifically we seek to accomplish the following:
Define and motivate a solution to the quantum-classical
correspondence problem independent from quantization, and discuss
the general insufficiency of prescriptive quantization,
particularly the Weyl quantization, to satisfy some axioms of
quantum mechanics. We achieve our purpose by addressing the the
time of arrival quantum-classical correspondence problem---the
problem of constructing time of arrival operators which is
consistent with the correspondence principle. The paper is
organized as follows. In Section-\ref{QCCP} we define what we
mean by the quantum-classical correspondence of dynamical
observables and introduce the concept of supraquantization; in
Section-\ref{TOAQCCP} we outline the solution to the time of
arrival QCC-problem; in Section-\ref{examples} we outline the
solution to the linear potential, the harmonic oscillator and the
quartic oscillator; in Section-\ref{compare} we compare Weyls'
quantization and the result of our treatment;  finally, in
Section-\ref{conclusion} we outline questions raised by the
present report. Our treatment is at the formal level. Its
rigorous version shall be given elsewhere.

\section{The Quantum-Classical Correspondence\\Problem}\label{QCCP}

The quantum-classical correspondence (QCC) problem  is generally
understood as the problem of deriving the quantum image of a
given classical system and from the quantum image recover
classical mechanics via an appropriate limiting procedure
involving the vanishing of the Planck's constant \cite{werner}.
The QCC problem has three aspects, the correspondence between
classical and quantum states, dynamics, and observables. In this
paper we limit ourselves to the quantum-classical correspondence
problem of dynamical observables. Thus we address the issue of
deriving the quantum image of a classical observable and
recovering the same classical observable from its quantum image.

Any solution to the QCC-problem for a given class of observables
consists of (i) a prescription of obtaining the corresponding
quantum observables, and (ii) a mapping of these to their
respective classical counterparts, i.e. an implimentation of the
correspondence principle \cite{werner}. A satisfactory solution
should solve the QCC-problem without conflict with the rest of
quantum mechanics. At present quantum observables are constructed
by the method of quantization, a prescriptive mapping of the
classical observables to quantum operators in a Hilbert space
\cite{quanti}; the classical limits of these are obtained by the
inversion of the prescribed mapping. However, this particular
solution has several unsatisfactory features as mentioned above,
most notorious is their general incompatibility with the required
values of commutators as it shall be demonstrated below for the
Weyl quantization for the Hamiltonian-time of arrival pair
\cite{grovan}.

If we seek a more satisfactory solution, then we have to drop
quantization and start elsewhere. But where?  The earlier work of
Mackey \cite{mackey} provides us with the motivation to seek
solution within quantum mechanics itself. We recall that the
quantization of free particle in one dimension is accomplished by
promoting its position and momentum $(q,p)$ into the operators
$(\opr{q},\opr{p})$ and their Poission bracket $\{q,p\}=1$ into
the commutator relation
$\left[\opr{q},\opr{p}\right]=i\hbar\opr{I}$, and the energy
$H=(2\mu)^{-1}p^2$ into the Hamiltonian operator
$\opr{H}=(2\mu)^{-1}\opr{p}^2$.  Mackey's work obviates these
quantization prescriptions by starting not from the classical
description but from the axioms of quantum mechanics and the
property of free space. Starting from the basic axiom that the
proposition for the location of the particle in different volume
elements are compatible and the fundamental homogeneity of free
space, one derives the position and the momentum operators
together with the canonical commutation relation they satisfy. On
the other hand, requiring Galilean invariance in the lattice of
propositions, one derives the free quantum Hamiltonian.

The free particle provides an excellent example of the existence
of more than one solution to the quantum-classical-correspondence
problem at the level of deriving the quantum image of a given set
of classical observables. While both solutions yield similar
results, they differ in many respects, thus distinct from one
another. The former presupposes classical mechanics, while the
latter upholds the autonomy of quantum mechanics; the former
introduces circularity when invoking the correspondence
principle, while the latter sanctions the correspondence
principle as a legitimate consequence of the acknowledged
preponderance of quantum mechanics over classical mechanics.
Should we make preference on one solution over the other? If we
required internal coherence and autonomy of quantum mechanics,
then the answer is yes. We are then obliged to accept the latter
in favor of the former. And seek solution in general to the
quantum-classical correspondence problem of its kind.

The former solution is an example of quantization.  The latter is
an example of what we shall refer to as {\it supraquantization} (
for the sake of differentiation). Supraquantization because it is
beyond quantization. Quantization is the derivation of the
quantum observable corresponding to a given classical observable
by means of a specified mapping of the c-valued observable to an
operator valued observable. Supraquantization is the derivation
of quantum observable corresponding to a given classical
observable without quantization. Quantization pressuposes the
axioms of classical mechanics. Supraquantization, on the other
hand,  presupposes the axioms of quantum mechanics including the
postulated properties of the physical universe and other
principles. Then, by definition, Mackey's construction of the
position and momentum operators together with their commutation
relation and the Hamiltonian is a supraquantization of their
classical counterparts. Mackey's construction required the axiom
of compatibility of the propositions for the location of the
particle,  the property of homogeneity of free space, and the
principle of Galilean invariance.

In both methods of obtaining quantum observables,  the classical
observable plays two different roles. In quantization, it is the
starting point; while in supraquantization, it is a boundary
condition. The correspondence principle requires that if a
quantum observable corresponds to a classical observable, then
the former should reduce to the latter in the limit of vanishing
$\hbar$. Then if supraquantization gives the correct quantum
observable, then that observable should approach its classical
limit. As a consequence of the role of the classical observable
as a boundary condition, supraquantization breaks the vicious
circle inherent in the quantization procedure.

In the following we demonstrate everything we have said about
supraquantization by addressing the time of arrival
quantum-classical-correspondence (TOA-QCC) problem.  We shall
limit ourselves in one dimension. The solution to the TOA-QCC
problem should contain the prescription of deriving the time of
arrival operator and the prescription of obtaining the classical
time of arrival. The second part has a standard solution and it
can now be given. Let $\kernel{\opr{A}}$ be the configuration
space kernel of the operator $\opr{A}$, then the classical limit
of $\opr{A}$ is given by
\begin{equation}\label{gentransition}
A(q,p)=\lim_{\hbar\to 0}2\pi \int\limits_{-\infty}^{\infty}\transker[\opr{A}]\,\exp\left(-i\frac{v\,p}{\hbar}\right) \,dv
\end{equation}
whenever the limit exists \cite{suttorp}. So when one knows how to obtain the kernel of the time operator, equation (\ref{gentransition}) provides the transition to the classical regime.

\section{The Time of Arrrival Quantum-Classical\\Correspondence Problem}\label{TOAQCCP}
\subsection{The Classical Time of Arrival}

Consider a particle with mass $\mu$ in one dimension whose
Hamiltonian is $H(q,p)$.  If the state of the particle at time
$t=0$ is given by the point $(q,p)$ in the phase space, what is
the time, $T_x$, that it will arrive at the point $q(t=T_x)=x$?
The solution to this problem is straightforward and is given by
\begin{equation}\label{classpass}
T_x(q,p)=\sign{p} \sqrt{\frac{\mu}{2}}\int_q^x \frac{dq'}{\sqrt{H(q,p)-V(q')}}
\end{equation}
whenever the integral exists and is real valued.  (For a derivation of equation (\ref{classpass}), see ref-\cite{leon}.) For a given energy $H(q,p)$, the region in the phase space in which equation (\ref{classpass}) exists and real valued is the classically accessible region to the particle. Because $T_x(q,p)$ is a time interval, it is canonically conjugate to the Hamiltonian, i.e. $\{H,T_x\}=1$. By virtue of $T_x$'s dependence on the phase space points $(q,p)$, $T_x$ is a dynamical observable.

It is our objective to derive equation (\ref{classpass}) from a quantum observable. But before we can address this problem, let us define the local time of arrival, $t_x (q,p)$, at a given point $x$ as the time of arrival at $x$ in some small neigborhood of $q$. The local time of arrival is technically the expansion of equation (\ref{classpass}) about the free time of arrival at $x$. For a given Hamiltonian $H=(2\mu)^{-1}p^2+V(q)$, the local time of arrival is given by
\begin{equation}\label{series}
t_{x} (q,p)=\sum_{k=0}^{\infty}(-1)^k\,T_k (q,p;x)
\end{equation}
where the $T_k (q,p;x)$'s are determined by the following
recurrence relation: $T_0 (q,p;x) = \mu p^{-1}(x-q)$ and
\begin{equation}\label{cowcow}
T_k (q,p;x)=\frac{\mu}{p}\int_{q}^{x} \frac{dV}{dq'}\frac{\partial T_{k-1}(q',p;x)}{\partial p}\,dq'
\end{equation}
for $k>0$. It can be shown that if $p\neq 0$ and if $V$ is
continuous at $q$, then there exists a neighborhood of $q$
determined by the neighborhood $\abs{V(q)-V(q')}<K_{\epsilon}\leq
(2\mu)^{-1}p^2$ such that for every $x$ in the said neighborhood
of $q$, $t_x (q,p)$ converges absolutely and uniformly to $T_x
(q,p)$.

Because $T_x(q,p)$ is defined in the entire accessible region of
the particle, we shall refer to it as the global time of arrival.
$t_x (q,p)$ converges to $T_x (q,p)$ only in a small neighborhood
so that $T_x(q,p)$ is the analytic extension of $t_x (q,p)$. In
the following sections we show that the local TOA, and thus the
global TOA by extension, can be determined completely from pure
quantum mechanical consideration. We shall limit though our
discussion for the time of arrival at the origin. Table-1 lists
the local and the global times of arrival at the origin for the
linear potential, the harmonic oscillator, and the quartic
oscillator. The local times of arrival are arrived recursively
using equation (\ref{cowcow}).

\begin{table}
\begin{center}
\begin{tabular}{|c|c|c|}\hline
Potential & Local Time of Arrival & Global Time of Arrival\\
$V(q)$ & $-t_{0}(q,p)$ & $-T_{0}(q,p)$\\
\hline\hline
$\lambda\, q$ &$ \displaystyle{\sum_{k=0}^{\infty}\frac{(-1)^k (2k)!}{2^k(k+1)!k!}\mu^{k+1} \lambda^k\frac{q^{k+1}}{p^{2k+1}}}$ & $\displaystyle{\frac{p}{\lambda}\left(\left(1+2\frac{\mu\lambda\,q}{p^2}\right)^{\frac{1}{2}}-1\right)}$\\
\hline
$\frac{1}{2}\omega^2 \mu q^2$ & $\displaystyle{\sum_{k=0}^{\infty}\frac{(-1)^k}{2k+1}\mu^{2k+1}\omega^{2k}\frac{q^{2k+1}}{p^{2k+1}}}$ & $\displaystyle{\frac{1}{\omega}\tan^{-1}\left(\frac{\mu\omega\,q}{p}\right)
}$\\
\hline
$\lambda\, q^4$ & $\displaystyle{\frac{\Gamma\left(\frac{3}{4}\right)}{8 \pi^{-\frac{1}{2}}} \sum_{k=0}^{\infty} \frac{(-2)^{k+1} \Gamma\left(-k-\frac{1}{4}\right)}{\Gamma\left(\frac{1}{2}-k\right)}\, \mu^{k+1}\lambda^{k} \frac{q^{4k+1}}{p^{2k+1}}}$ & $\displaystyle{\pm \sqrt{\frac{\mu}{2}}\!\! \int\limits_0^q\!\!\! \frac{dq'}{\sqrt{H(q,p) -\lambda {q'}^4}}
}$\\
\hline
\end{tabular}
\end{center}
\caption{The local and global times of arrival at the origin for the linear potential, harmonic oscillator and quartic oscillator.}
\end{table}

\subsection{Supraquantization of the Classical Time of Arrival}\label{supra}

Now we proceed in solving for the time of arrival
quantum-classical  correspondence by supraquantization. We shall
proceed formally. Consider the Hilbert space
$\mcal{H}=L^2(\Re,dq)$ and its particular rigging
$\Phi^{\times}\supset \mcal{H}\supset \Phi$ where $\Phi$ is the
fundamental space of all infinitely differentiable complex valued
functions with compact support in the configuration space $\Re$.
Given the self-adjoint Hamiltonian $\opr{H}$ which leaves $\Phi$
invariant under its action, we define its rigged Hilbert space
extension, $\opr{H}^{\times}$,  in the entire $\Phi^{\times}$ as
the operator
$\dirac{\opr{H}^{\times}\phi}{\varphi}=\dirac{\phi}{\opr{H}\varphi}$
for all $\phi\in \Phi^{\times}$ and $\varphi \in \Phi$. The
condition on the invariance of $\Phi$ under the Hamiltonian
restricts us to infinitely differentiable potentials which
includes entire analytic potentials. This is not a severe
restriction because we can always choose a different rigging of
the Hilbert space to accommodate other potentials. In the
following whenever refer to the Hamiltonian we mean its rigged
Hilbert space extension and shall continue to denote it by
$\opr{H}$.

For a given Hamiltonian $\opr{H}$ whose rigged Hilbert space
extension  is explicitly given by
\begin{equation} \label{Hamiltonian}
    \opr{H}\phi=-\frac{\hbar^2}{2\mu} \frac{d^2\phi}{dq^2}+V(q)\phi \;\;\;\mbox{for all} \;\;\phi \in \Phi^{\times},
\end{equation}
we assert that the corresponding time of arrival operator is
given by the operator $\opr{T}:\Phi\mapsto \Phi^{\times}$ whose
explicit action on $\Phi$ is
\begin{equation} \label{fretimo}
    \left(\opr{T}\varphi\right)\!(q)=\int_{-\infty}^{\infty}\kernel{\opr{T}}\varphi(q')\,dq' .
\end{equation}
in which the kernel must be symmetric, i.e.
$\kernel{\opr{T}}=\kernell{\opr{T}}^{*}$.  The solution then
depends on our ability to construct the time kernel,
$\kernel{\opr{T}}$,  for a given Hamiltonian. Once the time
kernel $\kernel{\opr{T}}$ has been solved, the classical time of
arrival operator can then be recovered by means of equation
(\ref{gentransition}), specifically
\begin{equation}\label{transition}
T_0(q,p)=\lim_{\hbar\to 0}2\pi \int\limits_{-\infty}^{\infty}\transker\,\exp\left(-i\frac{v\,p}{\hbar}\right) \,dv
\end{equation}
We will find below that under some  conditions equation
(\ref{transition}) reproduces exactly the local time of arrival
and, thus, the global one by extension.

But how do we determine the kernel $\kernel{\opr{T}}$ without
resorting  to quantization? We accomplish this in two steps.
First is by specifying the quantum evolution of $\opr{T}$. In
Heisenberg's representation, the time evolution of a quantum
observable $\opr{A}(t)=e^{i\opr{H}t}\opr{A}e^{-i\opr{H}t}$ is
governed by
$i\hbar\dot{\opr{A}}(t)=(\opr{A}(t)\opr{H}-\opr{H}\opr{A}(t))$,
in some appropriate domain.  If
$\opr{A}(t)\Phi\mapsto\Phi^{\times}$ for all $t$ and $\Phi$ is
invariant under $\opr{H}$, then the evolution equation can be
formally defined in the entire $\Phi$. Now if $\opr{T}$ is the
time of arrival operator, then it must evolve according to
$\dot{\opr{T}}(t)=-\opr{I}$. Imposing this condition we arrive at
the canonical commutation relation
\begin{equation} \label{condition}
    \Big(\big[\opr{H},\opr{T}\big]\!\varphi\Big)\!(q) = i\hbar\,\varphi(q) \,\,\,\mbox{for all}\,\,\phi\in\Phi
\end{equation}
satisfied by the Hamiltonian and the time  of arrival operator.
Equation (\ref{condition}) is the basic condition satisfied by
$\opr{T}$ but it is not sufficient to completely determine
$\opr{T}$.

The second step is by employing a transfer principle.  We
hypothesize that each element of a class of observables, such as
the time of arrival, shares a common set of properties with the
rest of its class such that when a particular property is
identified for a specific element of the class that property can
be transferred to the rest of the class without discrimination.

We exploit this in determining the kernel $\kernel{\opr{T}}$ by
solving  the simplest in the class of time of arrival
observables, the free particle. We start by recalling that the
free particle is Galilean invariant, a consequence of the
homogeneity of free space. It will not matter then where we place
the origin. This implies that the commutation relation
(\ref{condition}) holds independent of the choice of origin.
Because of this and because the free Hamiltonian is Galilean
invariant, we require that the time kernel for the free particle
must itself be Galilean invariant. Specifically if $t_a$ is
translation by $a$, i.e. $t_a(q)=q+a$ and if $\kernel{\opr{T}}$
is the free particle kernel, then the translated free time of
arrival operator $\opr{T}_a=\int dq\,\tkernel{\opr{T}}$ must
still satisfy equation (\ref{condition}). In addition to
translational invariance, $\kernel{\opr{T}}$ must be symmetric,
and it must be chosen such that equation (\ref{condition}) is
satisfied given the free Hamiltonian
$\opr{H}\phi=-\hbar^2(2\mu)^{-1}\phi''$, and it must reproduce
the free time of arrival at the origin via equation
(\ref{transition}).  A solution satisfying all these requirements
is given by
\begin{equation}\label{freefree}
\kernel{\opr{T}}=\frac{\mu}{i\,4\hbar} (q+q')\, \mbox{sgn}(q-q').
\end{equation}
We note though that (\ref{freefree}) is not unique. The kernel $\hbar^{-1}\mu \left|a-a'\right|$ is dimensionally consistent with (\ref{freefree}) and it is Galilean invariant and it commutes with the free Hamiltonian in the entire $\Phi$. Moreover it vanishes in the classical limit. Then real factors of it can be added to (\ref{freefree}) without sacrificing any of the required properties of the free particle kernel. However, $\hbar^{-1}\mu \left|a-a'\right|$ arises only because of Galilean invariance which is an exclusive property of the free particle.  Since we are aiming at exploiting the assumed transfer principle, we can not carry it over to the rest of its class. We shall have more to say about this latter.

Having solved the free particle kernel, we proceed in
implementing the  transfer principle. We hypothesize that all
time kernels assume the same form. Thus, from equation
(\ref{freefree}), we assume that the time kernel is given by
\begin{equation} \label{timekernel}
    \kernel{\opr{T}}=\frac{\mu}{i\,\hbar}\; T(q,q')\, \mbox{sgn}(q-q')
\end{equation}
where $T(q,q')$ depends on the given Hamiltonian. Inferring from
the free particle time kernel, we require that $T(q,q')$ be real
valued,  symmetric, $T(q,q')=T(q',q)$, and analytic. We determine
$T(q,q')$ by imposing condition (\ref{condition}) on $\opr{T}$.
Substituting equation (\ref{timekernel}) back into the left hand
side of equation (\ref{condition}) and performing two successive
integration by parts, we find that $T(q,q')$ must satisfy the
partial differential equation, which we shall refer to as the
time kernel equation,
\begin{equation} \label{kerneldiff}
 -\frac{\hbar^2}{2\mu}\frac{\partial^2 T(q,q')}{\partial q^2}+\frac{\hbar^2}{2\mu} \frac{\partial^2 T(q,q')}{\partial {q'}^2}+ \left(V(q)-V(q')\right)T(q,q') =0
\end{equation}
and the boundary condition
\begin{equation}\label{origbound}
    \frac{d T(q,q)}{dq} + \frac{\partial T(q',q')}{\partial q} +    \frac{\partial T(q,q)}{\partial q'}=1.
\end{equation}
for all $q,q'\in\Re$. The boundary condition (\ref{origbound}) defines a family of operators canonically conjugate to the extended Hamiltonian in the sense required by equation (\ref{condition}). This is a reflection of the fact that there are numerous operators that are canonically conjugate to a given Hamiltonian. We fix the boundary condition by refering to the free particle. The $T(q,q')$ of the free particle is $\frac{1}{4}(q+q')$. By inspection $T(q,q')$ satisfies both (\ref{kerneldiff}) and (\ref{origbound}). Again invoking the assumed transfer principle, we arrive at the following boundary conditions for the time of arrival at the origin,
\begin{equation}\label{equiv}
T(q,q)=\frac{q}{2},\;\;\;\;T(q,-q)=0.
\end{equation}
We claim that equations (\ref{kerneldiff}) and (\ref{equiv})
constitute the supraquantization of the local time of arrival
consistent with the correspondence principle. We will demonstrate
this claim below. (Other methods of constructing time of arrival
operator without quantization are given in references \cite{muga}
and \cite{leon}. Their motivation, however, is different from
ours.)

We remark that our ability to extract probability distributions from the time of arrival operators defined by equation (\ref{fretimo}) depends on whether the operators (\ref{fretimo}) have nontrivial Hilbert space projections. If their restrictions on the Hilbert space exists, then they will be generally unbounded and non-self-adjoint. Then they will be classified as POV-observables. Finding the spectral resolution of the time of arrivals is a non-trivial problem and it shall not be addressed here. This will be the topic of a more extensive and indepth treatment of the time of arrival quantum classical corresponence problem.

\section{Explicit Examples}\label{examples}

Table-2 summarizes the solution to the  time kernel equation for
the linear potential, the harmonic oscillator and the quartic
oscillator. We have arrived at the solutions by changing
variables from $(q,q')$ to  $(u=q+q',v=q-q')$. The differential
equation (\ref{kerneldiff}) and the boundary condition then
assume the form
\begin{equation}\label{newdequation}
-2\frac{\hbar^2}{\mu}\frac{\partial^2 T}{\partial u\,\,\partial v} + \left(V\left(\frac{u+v}{2}\right)-V\left(\frac{u-v}{2}\right)\right) T(u,v)=0
\end{equation}
\begin{equation}\label{newboundary}
T(u,0)=\frac{u}{4},\;\;\;T(0,v)=0
\end{equation}
We have sought an analytic solution  in powers of $u$ and $v$,
i.e. $T(u,v)=\sum_{m,n\geq 0}\alpha_{m,n} \,u^m v^n$ where the
$\alpha_{m,n}$'s are constants determined by equations
(\ref{newdequation}) and (\ref{newboundary}) for a given
potential. Once the solution to equation (\ref{newdequation}) for
a given potential has been found, we transform back to $(q,q')$
to get the results in Table-2.

\begin{table}
\begin{center}
\begin{tabular}{|c|c|}\hline
Potential & Solution to the Time Kernel Equation\\
$V(q)$ & $T(q,q')$\\
\hline\hline
$\lambda\, q$ &$ \displaystyle{\frac{1}{4}\sum_{k=0}^{\infty}\frac{1}{k!(k+1)!} \left(\frac{\mu\lambda}{4\hbar^2}\right)^k \,(q+q')^{k+1}(q-q')^{2k}}$\\
\hline
$\frac{1}{2}\omega^2 \mu q^2$ & $\displaystyle{\frac{1}{4}\sum_{k=0}^{\infty} \frac{1}{(2k+1)!}\left(\frac{\mu\omega}{2\hbar}\right)^{2k} (q+q')^{2k+1} (q-q')^{2k}}$\\
\hline
$\lambda\, q^4$ & $\displaystyle{\frac{1}{4}\sum_{m=0}^{\infty}\sum_{n=2m}^{\infty}\Delta_{m,n} \left(\frac{\mu\lambda}{8 \hbar^2}\right)^{n-m} \,(q+q')^{4n+1-6m}\,(q-q')^{2n}}$\\
\hline
\end{tabular}
\end{center}
\caption{The solution to the time kernel  equation for the linear
potential, harmonic oscillator and quartic oscillator.}
\end{table}

Before we proceed let us define the following transform of the time kernel,
\begin{equation}\label{transkerker}
\mathcal{T}_{\hbar}(q,p)=2\pi\!\! \int_{-\infty}^{\infty}\!\!\!\transker\,\exp\left(-i\frac{v\,p}{\hbar}\right) \,dv.
\end{equation}
$\mathcal{T}_{\hbar}(q,p)$ is real valued and odd with respect to $p$. First let us consider the linear potential and the harmonic oscillator. Both systems have linear classical equations of motion, or simply linear systems. Calculating $\mathcal{T}_{\hbar}(q,p)$ for these systems on using the identity
\begin{equation}\label{identity}
\int_{-\infty}^{\infty}\sigma^{m-1}\mbox{sgn}(\sigma)\,\exp(-ix\sigma)\,d\sigma= \frac{(m-1)!} {i^m\pi} x^{-m},
\end{equation}
it is straightforward to show that it exactly reproduces  the
local time of arrival at the origin as given by Table-1, i.e.
\begin{equation}\label{linharm}
\mathcal{T}_{\hbar}(q,p)=t_0 (q,p)
\end{equation}
for both systems.

We turn to the quartic oscillator. In comparison with the previous two examples, it is nonlinear in the sense that it has nonlinear classical equation of motion. The $\Delta_{m,n}$'s in the solution are constants satisfying the recurrence relation $(4n+1-6m)n \Delta_{m,n}=\Delta_{m,n-1}+\Delta_{m-1,n-2}$ with $\Delta_{0,0}=1$, for $n\geq 2m$, $m,n\geq 0$. For our present purposes we don't gain anything by writing down explicitly the $\Delta_{m,n}$'s. It is sufficient to give the value of $\Delta_{0,n}=\left(4^{n+2}n!\Gamma\left(\frac{3}{4}\right)\right)^{-1}(-1)^n \Gamma\left(-\frac{1}{4}-n\right)$. Calculating for its $\mathcal{T}_{\hbar}(q,p)$, we find
\begin{eqnarray}\label{anharm}
\mathcal{T}_{\hbar}(q,p)&=&2\mu \sum_{m=0}^{\infty}\hbar^{2m}\!\!\! \sum_{n=2m}^{\infty}(-1)^n \,\Delta_{m,n}(2n)!  \left(\frac{\mu\lambda}{8}\right)^{n-m}\frac{(2q)^{4n+1-6m}}{p^{2n+1}}\nonumber\\
&=&\frac{\Gamma(3/4)}{8 \pi^{-\frac{1}{2}}} \sum_{n=0}^{\infty} \frac{(-2)^{k+1}\Gamma\left(-\frac{1}{4}-k\right)}{\Gamma\left(\frac{1}{2}-k\right)} \mu^{n+1}\lambda^{n} \frac{q^{4n+1}}{p^{2n+1}}+\mathcal{O}(\hbar^2)\nonumber\\
&=&t_0 (q,p) + \mathcal{O}(\hbar^2)
\end{eqnarray}
We see that equation (\ref{anharm}) reduces only to the
classical  time of arrival in the limit of vanishing $\hbar$ or
infinitesimal $\hbar$.

The above examples are special cases of the following general result for entire analytic potentials,
\begin{equation}
\mathcal{T}_{\hbar}(q,p)= \left\{\begin{array}
                        {r@{\quad:\quad}l}
                        t_0 (q,p)\!\!\! &\!\!\! \mbox{linear systems}\\
                        t_0 (q,p) + \mathcal{O}(\hbar^2)\!\!\! &\!\!\! \mbox{non-linear sytems}
                        \end{array}\right.
\end{equation}
For these systems $\lim_{\hbar\to 0}\mathcal{T}_{\hbar}(q,p)=t_0 (q,p)$, and to $T_0 (q,p)$ by extension, in keeping with the correspondence principle. We can then claim that, at the formal level, equations (\ref{fretimo}) and (\ref{transition}) constitute a solution to the time of arrival quantum-classical correspondence problem without quantization.

\section{The Insufficiency of Weyl Quantization}\label{compare}

Weyl quantization is a prescription which maps a given  classical
dynamical observable into a quantum observable \cite{weyl}. It is
based on some symmetric ordering of non-commuting observables.
Let $f(q,p)$ be an observable defined on the phase space.  Weyl
quantization maps $f(q,p)$ into the operator $\opr{W}_{[f]}$.
$\opr{W}_{[f]}$ acts on the configuration space functions $\psi$
according to
\begin{equation}
\left(\opr{W}_{[f]}\psi\right)\!(q)=\int_{-\infty}^{\infty} \kernel{\opr{W}_{[f]}}\psi(q')\,dq'
\end{equation}
where the kernel is
\begin{equation}\label{kerker}
\kernel{\opr{W}_{[f]}}=\frac{1}{2\pi \hbar}\int_{-\infty}^{\infty} f\!\left(\frac{q+q'}{2},p\right) \exp\left[\frac{i}{\hbar}(q-q')p\right]\, dp
\end{equation}
Weyl quantization has the remarkable property that it is
translationally  invariant. Given the kernel of $\opr{W}_{[f]}$
one recovers the classical observable by a inversion of the Weyl
transform,
\begin{equation}\label{class}
f(q,p)=2\pi \int_{-\infty}^{\infty} \transker[\opr{W}_{[f]}]\exp\left(-\frac{i}{\hbar}v\,p\right)\,dv
\end{equation}
Equations (\ref{kerker}) and (\ref{class}) establish a one-to-one
correspondence  between classical observables and their quantum
counterparts via Weyl's prescription.

Equations (\ref{timekernel}) and (\ref{kerker}) prescribe the
kernel  for the local time of arrival operator. Now we compare
their results. Using the identity
\begin{equation}
\int_{-\infty}^{\infty}x^{-m} \exp(i x \sigma)\, dx=\frac{i^{m}\pi}{(m-1)!}\sigma^{m-1}\, \mbox{sgn}(\sigma)
\end{equation}
to solve for the Weyl kernels of the local  time of arrivals in
Table-1, we find the following: For the linear and harmonic
oscillator, we have the equality
$\kernel{\opr{T}}=\kernel{\opr{W}_{[t_0]}}$.  Thus for these
cases the results of quantization and supraquantization are the
same. However, for the quartic oscillator, we find the inequality
$\kernel{\opr{T}}\neq\kernel{\opr{W}_{[t_0]}}$.  In fact one can
show that the leading term in the $\kernel{\opr{T}}$ of the
quartic oscillator is the Weyl kernel of the same system. Because
of this and equation (\ref{fretimo}) is canonically conjugate to
the Hamiltonian in the sense of equation (\ref{condition}), the
inequality $\kernel{\opr{T}}\neq\kernel{\opr{W}_{[t_0]}}$ implies
that the Weyl quantization of the local time of arrival for the
quartic oscillator fails to be canonically conjugate with the
Hamiltonian, even though the local TOA and the classical
Hamiltonian are.  This is another demonstration of the well known
limitation of the Weyl quantization, and prescriptive
quantizations in general, in satisfying required commutator
values.

Our examples above demonstrate our general result for entire
analytic potentials that only for linear systems that Weyl
quantization is consistent and that for non-linear systems Weyl
quantization only yields the leading term to the time kernel:
\begin{equation}\label{solsol}
\kernel{\opr{T}}=\left\{ \begin{array}
                        {r@{\quad:\quad}l}
                        \kernel{\opr{W}_{[t_0]}} & \mbox{linear systems}\\
                        \kernel{\opr{W}_{[t_0]}}+\Delta(q,q')& \mbox{non-linear sytems}
                        \end{array}\right.
\end{equation}
where $\Delta(q,q')$ is the correction term introduced by the non-linearity of the system.

\section{Conclusion}\label{conclusion}
In this paper we have raised the issue and the problem of
constructing quantum observables without quantization. Our
results demonstrate the capability of supraquantization to yield
acceptable solutions where quantization fails. We recognize
though that we have raised more questions in attempting to answer
one. For example, is the present set of axioms of quantum
mechanics sufficient to derive all of classical dynamical
observables?  If we strongly assert the autonomy and internal
coherence of quantum mechanics, then it may be necessary to
modify the present axioms to give satisfactory solution to the
quantum-classical correspondence problem. An example of such
modification is in the expansion of the definition of quantum
observables to include positive operator valued measures
\cite{busch}. Exclusion of these would lead the exclusion of
numerous classical observables whose quantum image are
non-self-adjoint operators. A further expansion of quantum
observables would include subnormal operators. Such operators
would be the appropriate quantum image of the reciprocal of
classical observables whose quantum image are non-invertible,
e.g. the inverse momentum for periodic boundary conditions in a
segment of the real line.

Also the question of the uniqueness  of the  solution to the
quantum-classical correspondence problem arises. As evidenced by
the free particle the solution may not be unique. (See
Section-\ref{supra}.) However, the non-uniqueness arises from the
manner in which we take the classical limit, particularly the
manner in which $\hbar$ vanishes in the classical regime. The
$\mathcal{T}_{\hbar}$ transform of $\left|q-q'\right|$, equation
(\ref{transkerker}), is to the order $\mathcal{O}(\hbar)$ so that
in the limit of vanishing $\hbar$ it does not contribute
anything. Since $\hbar^{-1}\mu \left|q-q'\right|$ commutes with
the free Hamiltonian in $\Phi$ and it is Galilean invariant,
equation (\ref{freefree}) is arbitrary up to an additive real
multiple of $\hbar^{-1}\mu \left|q-q'\right|$.  However, if we
take classical mechanics to be the limiting theory of quantum
mechanics where $\hbar$ is infinitesimal, i.e. only second and
higher powers of $\hbar$ vanish, then
$\hbar^{-1}\mu\left|q-q'\right|$ will introduce unwanted
correction to the classical time of arrival. If we adhere
strictly to the correspondence principle as implemented by the
manner in which $\hbar$ disappear in the classical domain,
$\hbar=\delta$, then $\hbar^{-1}\mu \left|q-q'\right|$ has to be
dropped. This shows that supraquantization may not be independent
from the manner in which classical observables are recovered from
quantum observables.

Our answers above are but cursory and the questions still require
greater depth than where we are at  present.  Moreover, there are
many other questions aside from the above two (For example, how do
we determine which set of axioms goes with a given problem? Is
there a general framework upon which the solution to the
QCC-problem without quantization can be built upon?), but right
now we can only hope to address them in the future.

\section*{Acknowledgement}
The author is grateful to  Prof. Kilin for the  invitation to
participate in the conference and appreciates the warm
hospitality of E. Tolkacheva and D. Horoshko. The author
acknowldges the support given by the Commission on Higher
Education which made his travel possible.

\end{document}